\documentclass{aa}
\usepackage{natbib}
\usepackage{epsfig}
\usepackage{wasysym}
\begin{document}
%%%%%%%%%%%%%%%%%%%%%%%%%%%%%%%%%%%%%%%%%%%%%%%%%%%%%%%%%%%%%%%%%%%%%%%%%%%
\title{Bayesian detection of planetary transits}

   \subtitle{A modified version of the Gregory-Loredo method for
   Bayesian periodic signal detection} 

   \author{S. Aigrain\inst{}\thanks{Now at IoA -- Madingley Road, Cambridge
                                    CB3 0HA, United Kingdom}
          \and
          F. Favata\inst{}
          }

   \offprints{S. Aigrain}

   \institute{Astrophysics Division -- 
              Space Science Department of ESA, ESTEC, 
              Postbus 299, NL-2200 AG Noordwijk, The Netherlands\\
              \email{suz@ast.cam.ac.uk, Fabio.Favata@rssd.esa.int}
             }

   \date{Received \ldots; accepted \ldots}

   \abstract{ The detection of planetary transits in stellar
     photometric light-curves is poised to become the main method for
     finding substantial numbers of terrestrial planets. The
     French-European mission COROT (foreseen for launch in 2005) will perform
     the first search on a limited number of stars, and larger
     missions \emph{Eddington} (from ESA) and \emph{Kepler} (from
     NASA) are planned for launch in 2007. Transit
     signals from terrestrial planets are small ($\Delta F/F \simeq
     10^{-4}$), short ($\Delta t\simeq 10$ hours) dips, which repeat with 
     periodicity of a few months, in time series lasting up to a few years. 
     The reliable and automated detection of such signals in
     large numbers of light curves affected by different sources
     of noise is a statistical and computational challenge. 
     We present a novel algorithm based on a Bayesian
     approach. The algorithm is based on the Gregory-Loredo method
     originally developed for the detection of pulsars in X-ray
     data. In the present paper the algorithm is presented, and its
     performance on simulated data sets dominated by photon noise is
     explored. In an upcoming paper the influence of additional noise
     sources (such as stellar activity) will be discussed.
   \keywords{Planetary systems --
                Occultations --
                Methods: Data analysis
               }
   }

   \maketitle

\section{Introduction}
\label{intro}

The search for rocky, terrestrial planets around other stars is a key
research topic in astrophysics for the next decade. Following the
first exo-planet detection around a sun-like star \citep{mq95}, gaseous 
giants around other solar-type stars have been shown to be relatively common 
\citep{bmf01}. The mass function of the current crop of extra-solar planets 
grows rapidly toward the lower masses \citep{bmf01}, showing that low-mass 
planets must be common. However, the radial velocity technique, which has 
resulted in the detection of the exo-planets detected so far, is limited to 
planetary masses somewhat smaller than Saturn, and cannot reach the domain of 
terrestrial planets. This is due to astrophysical effects, such as 
microturbulence in the star's atmosphere, rather than instrumental limitations.

The most promising approach for the detection of (significant numbers)
of terrestrial planets around stars other than the Sun appears to be
the search for planetary transits, i.e.\ dips in the light curve of
the parent stars caused by the planet transiting in front of the
stellar disk. The flux dip caused by the transit is also small,
$\Delta F/F = (R_{\mathrm{p}}/R_*)^2$, which for the transit of an Earth-Sun
system gives $\Delta F/F = 10^{-4}$. This is well below the scintillation
noise caused by the Earth's atmosphere \citep[see]{f+00}, so that high-accuracy
space-based photometry will be needed for the detection of such
events. The probability of occurrence of a transit depends on the
inclination of the planetary orbit relative to the line of sight
(which must be close to $i = 90$ degrees), and is relatively small (for a
set of randomly oriented Sun-Earth systems $p \simeq 0.5\%$), so that
searches for planetary transits must be based on observation of large
samples of target stars. A typical transit duration will be of order
$\Delta t \simeq 10$ hours, and the transit periodicity will be the same
as the orbital period of the planet, typically several months.

A number of space missions wholly or partially dedicated to the search
for planetary transits are either in development or in the planning
stage. The CNES/European satellite COROT is planned for launch in
2005, while the ESA mission \emph{Eddington} and the NASA mission
\emph{Kepler} are planned for launch in 2007. Given the
intrinsically statistical nature of planetary transit searches, these
missions will acquire large number of stellar light curves, ranging
from thousands for COROT to hundreds of thousands for \emph{Eddington}
and \emph{Kepler}. Also, some smaller searches are being conducted for
limited time periods (and concentrating on larger planets) using e.g.\ 
HST (\citealp{gbg+2000}) or ground-based telescopes (e.g.\
\citealp{ddk+2000}). 

The analysis of data from such searches, and in particular the
detection of transits with a high degree of certainty and a low false alarm
rate, is a challenging task. The transit signal is weak ($\Delta F/F =
10^{-4}$), and concentrated in a small fraction of the total signal:
for a habitable planet orbiting a K5V star the orbital period will be
roughly 4 months, so that for a 1 year light curve three events will be
present. As each transit lasts $\approx$ 10 hours, the transit signal is
present in only $\approx 0.3\%$ of the total light curve.  In the Euclidean 
regime, the number of stars in a given field increases toward fainter 
magnitudes by a factor of $\approx 4$ per
magnitude. This is the case for the range of magnitudes and the low Galactic 
target latitudes of interest for currently planned missions . 
Therefore, most of the detected planets will be in the light curves of the 
fainter (and thus statistically noisier) stars, impying the need for 
effective robust data analysis algorithms able to reliably detect transits 
``hidden in the noise''. At the same
time, the large number of light curves which will need to be analyzed,
each with a large number of points (of order $10\,000$ points for a
year of data) makes the use of efficient algorithms necessary, and
rules out brute force approaches.

Some ground- (\citealp{ddk+2000}) and HST-based (\citealp{gbg+2000})
transit searches, which deal with relatively small numbers of light
curves, use a detection approach based on comparing large numbers of model 
transits to the light curves and minimising a $\chi^2$ statistic
(or a linear statistic in the case of Doyle). These approaches are 
computationally very intensive, and thus may be unsuitable for the routine 
processing of the large number of light curves which will be produced by 
upcoming space missions.

As an alternative, transit detection algorithms based on Bayesian
methods have recently been the subject of some attention. They have the
advantage of combining computational efficiency with flexibility.
While a global statistic can be used for the detection, information is 
directly available to reconstruct the detected signal if wanted, therefore 
providing a tool to discriminate between planetary transits and other types 
of periodic signals (\citealp{dbd2001}), as well as directly measuring 
additional planetary characteristics such as the planet's radius.

In the present paper we present a novel algorithm for the detection of
planetary transits based on the method developed by \citet{gl92b}
(hereafter referred to as GL method) for the search of pulsed emission
from pulsars in X-ray data. While the algorithm was developed to be
``general purpose'', we have tuned it with the parameters of
the upcoming \emph{Eddington} planet finding mission in mind. The present
paper discusses the characteristics of the algorithm on the basis of
extensive simulations for the case in which the light curve is
dominated by photon noise. Its performance in the case in which
stellar activity is the dominating noise source will be the subject of
a future paper.

Bayesian algorithms for the detection of planetary transits are also
being developed in the context of the COROT mission. In particular, an
approach based on expansion of the light curve into a truncated
Fourier series is being investigated (\citealp{ddb2001}). Perfoming the 
detection in the Fourier domain can make the algorithm computationally 
sensitive to data gaps and sampling rates. Here we explore a more robust
\emph{direct space} approach.

The GL (\citealp{gl92b}) method, was initially developed for the detection of 
X-ray pulsars (where Poisson statistics dominate) and later extended to the 
Gaussian noise case (\citealp{gre99}). At the flux levels of interest for the 
transit searches for \emph{Eddington}, the photon shot noise per detection 
element (which is Poissonian) can be very well represented by Gaussian noise 
(see Sect.~\ref{trans}). The original formulation of the GL algorithm is 
well-suited to the detection of periodic signals of unknown shape. However, 
in the planetary transit problem we have strong prior information about the 
transit shape. In this paper we modify the GL algorithm to perform more 
optimally for planetary transit detection. We do this by allowing one of the 
bins to have a variable width, to represent the out of transit constant signal 
level. This formulation also permits the phase of the transits to be 
identified, a task the original GL method is not suited for (see 
Sect.~\ref{bay}). The fitted parameters are the period, duration and phase of 
the transit. The shape of the transit can then be reconstructed from the 
phase-folded light curve.

The simulated light curves are described in Sect.~\ref{lc}. The algorithm
is outlined in Sect.~\ref{algo} and compared with the original GL algorithm in 
Sect.~\ref{comp}. Sect.~\ref{perf} describes the evaluation of the algorithm's 
performance by determining the number of false alarms and missed detections in 
a large sample of simulated light curves with and without transits.  
Conclusions and options for future work are presented in Sect.~\ref{concl}.

%__________________________________________________________________

\section{The light curves}
\label{lc}

\subsection{Transits}
\label{trans}

Given the presence of limb darkening in stellar photospheres,
planetary transits are not perfectly ``flat bottomed'' (nor are they,
strictly speaking, truly grey). To simulate transits in a realistic
way, the Universal Transit Modeler (UTM) software written by H.J.~Deeg 
(\citealp{dee99}) was used. Limb darkening coefficients were taken from
\citet{ham93}. Two types of transits were simulated,
one representing a Jupiter-type planet in a short orbit around a
Sun-like star and another representing a Earth-like planet in an
habitable orbit around a K5V star.

The input characteristics of the system for the Jovian transit were:

\begin{itemize}
 
\item Time step $t_{\mathrm{unit}} = 15$ minutes.
 
\item Radius of star $R_* = R_\odot$.

\item Luminosity of star $L_* = L_\odot$.

\item Radius of planet $R_{\mathrm{p}} = R_{\jupiter}$. The ratio of the 
  duration of the ingress / egress, to that of the `flat bottom' of the 
  transit (affected only by limb darkening) is roughly 
  $ 2\mathrm{R_p} / (\mathrm{R}_{\star}-\mathrm{R_p})$, in this case 
  $\simeq 0.22$.

\item Period of transit $P = 2880 \times t_{\mathrm{unit}}$ 
  ($1$ month).
  This is at the short end of the range of periods of interest for
  \emph{Eddington}, but extrapolation to longer periods is to some
  extent possible (when activity is not included) by acting on the
  number of transits in the light curve.

\item The distance star-planet, which is used to determine the
  transit duration, was varied between 
  $d = 15.3 \times R_{\odot}$
  (resulting in a transit duration of 15 hours) and 
  $d = 45.9 \times 
  R_{\odot}$ (5 hours). (N.B.: for the same period different distances would 
  correspond to different planetary to stellar mass ratios).

\item The duration of light curve was varied between $D = 3 \times
  P$ and $D = 5 \times P$.

\item The phase of the transit was randomly varied in the different
  simulations. The posital phase (between 0 and 1) is used in the
  course of the present paper.

\end{itemize}

Light curves were normalized to the photon count level expected for a
star of a given $V$ magnitude (between 7 and 17 typically), based on
the throughput expected for the baseline \emph{Eddington} mission
design (\citealp{f+00}), i.e.\ a collecting area of $0.6$ m$^2$ and
a total system throughput of $70\%$. With these instrument parameters a
$V=21.5$ G2V star will yield $\simeq 50$ detected photons/sec.
Gaussian noise was then added with variance defined by the number of 
detected photons per pixel.

%__________________________________________________________________

\section{The algorithm}
\label{algo}

\subsection{A Bayesian method}
\label{bay}

The method employed consists in calculating the likelihood of the data
given a certain number of parameters, varying the parameters over a
given range and identifying the value of each parameter whose
probability is maximized according to Bayes' theorem:

\begin{equation}
\label{eqprob}
   p(\theta|data,I) = p(\theta|I) \times \frac{p(data|\theta,I)}{p(data|I)}
\end{equation}
where :
\begin{itemize}
  \item $\theta$ is a set of parameter values (i.e. a hypothesis).
  \item $data$ is the dataset.
  \item $I$ represents information about the ensemble of hypotheses considered
   i.e. the type of model used and knowledge about the other models. 
   For the remainder of this section
   $I$ will be implicit in likelihood expressions.
  \item $p(data|I)$ is a prior for the type of model used.
  \item $p(\theta|I)$ is the combined prior for the parameters.
\end{itemize}

An excellent description of the theory on which the present algorithm
is based is given in \citet{gl92b}. In the present paper we will
give a brief outline of the calculations, detailing only those aspects
in which our work differs from the discussion of \citet{gl92b}. As a
starting point we constructed an algorithm following exactly
the GL prescription, and we tested it on sets of 10 simulated light
curves containing transits with varying characteristics. This
benchmark was later used to ensure that the modifications in our
algorithm were indeed improvements. 

\begin{figure}[htb]
  \centering
  \epsfig{file=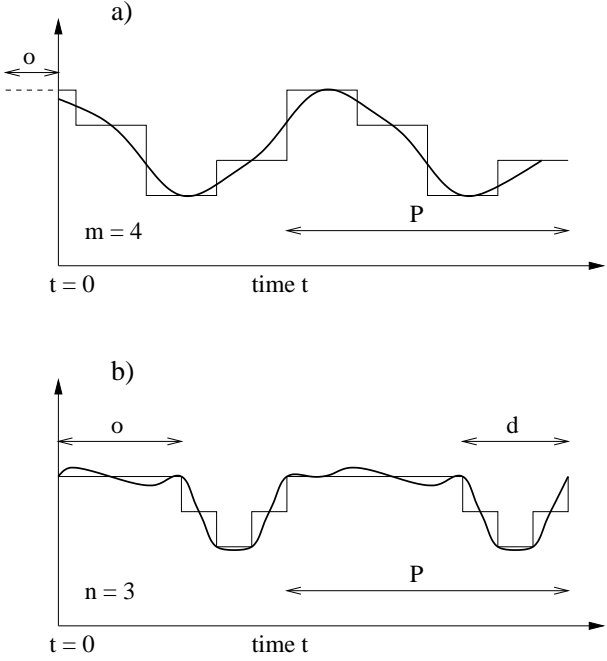,width=8.0cm, angle=0} 
  \caption{Schematic illustration of the type of model and parameters used. 
    a) GL method with $m=4$. b) Modified method with $n=3$. }
  \label{fig:params}
\end{figure}

The GL algorithm employs a family of stepwise models to describe the
periodic signal plus background. Each member of the family resembles a
histogram, with $m$ equal sized bins per period $P$. The family members are
distinguished by the value of $m$ which is varied in the range from 2 to
some upper limit (typically 15 for X-ray pulsar detection work). Such a
model is capable of approximating a light curve of essentially arbitrary
shape, which is desirable for detecting periodic signals of unknown shape,
in contrast to the current planetary transit problem, for which the shape is 
known a priori. GL also employs a phase parameter $\phi$. If the time offset 
$o$ is defined as the  time elapsed between the start of the first bin and 
the start of the data, $\phi = 2\pi (o/P)$. The parameters fitted by the GL 
method are then $P$, $m$, $\phi$ (the flux level in each bin of the model is 
marginalised over). In the case of planetary transits, it is not desirable to 
let the model vary outside the transit. We therefore have a slightly different 
type of model. The number of steps in the step function is $n+1$. Bin $0$ is 
the `out of transit' bin and lasts for a large fraction of any given period, 
and bins $1$ to $n$ are `in transit', each lasting $d/n$ where $d$ is the 
duration of the model transit. We have also adopted a different definition for 
the time offset, as we have a significant event -- the transit -- which we can 
use to determine the start of a new period. Defining the time offset $o$ as 
the time from the start of the data to the start of the next transit, the 
phase is then related to the offset in the same way as before. The parameters 
required are now $P$, $\phi$ and $d$. These parameter definitions are 
illustrated for both methods in Fig.~\ref{fig:params}

As there is no feature in a step function of unconstrained shape with equal 
duration steps which can mark the beginning of a period in the data, the 
concept of phase is not well defined. Any bin in the step function could be 
the first. Thus we do not expect the GL method to enable phase determination 
directly. Only after reconstruction of the entire light curve could the 
position of the transit be pin-pointed relative to the start of the data. By 
introducing a transit feature in the model, the phase is built into the model 
function and we expect it to be detected effectively by the modified algorithm.

Models with a lower number $n$ of `in transit' bins will incur a lower Occam 
penalty factor, as emphasised in \citet{gl92b}. In general, $n$ should be 
chosen to be the lowest value possible. For pure detection purposes, given 
that transits are relatively simple events, $n=1$ should suffice. For transit 
reconstruction purposes, a higher value can be used. 

Despite the modifications we made to the GL models, we followed the method 
outlined in \citet{gre99} to calculate the likelihoods.

\subsection{Likelihood calculation}
\label{likcalc}

The likelihood is initially calculated for a given set of parameters
$P$ (period), $d$ (duration), $o$ (offset). For convenience the results 
were sometimes 
expressed in terms of posital phase: ${\mathrm{ph}} = \phi / 2 \pi = o / P$.

Due to the different type of model function used, Eq.~(6) in
\citet{gre99}, which describes the assigmnent of a bin number $j$ to each 
data point $y_i$ taken at time $t_i$, was replaced by the following:

\begin{equation}
  \label{eqjyes}
  j(t_i) = 
    \left\{ 
      \begin{array}{lcl}
        t_{\mathrm{mod}} & : & \textrm{if } 0 < t_{\mathrm{mod}} \leq n \\
        0 & : & \textrm{otherwise}
      \end{array}
    \right.
\end{equation}

\noindent where:
\begin{equation}
  \label{eqtmod}
  t_{\mathrm{mod}} = \mathrm{int} \left( 
    \frac{ (\mathnormal{t_i + P - o}) ~\mathrm{mod}~ (\mathnormal{P})}
         {\mathnormal{d/n}}
    + 1 \right)
\end{equation}
\noindent $n$ is the number of bins per transit, 
$\mathrm{int}(\mathnormal{x})$ is the nearest integer lower than or equal to
$x$ and $(a) ~\mathrm{mod}~ (\mathnormal{b})$ is the remainder of $a$ divided 
by $b$.

At time $t_{i}$, the observed flux count $y_{i}$ can be written as
$y_{i}=y(t_{i})+e_{i}$ where $y(t_{i})$ is the value predicted by the
model for time $t_i$ and $e_{i}$ is a noise component. The noise is assumed 
to have a Gaussian distribution (see \citealp{gre99} and references therein) 
with variance $\sigma_{i}^{2}$. In the present case it is appropriate and 
clearer to use the same value of $\sigma$ for all data points\footnote{In 
\citet{gre99} a noise parameter $b$ is introduced to account for incomplete 
knowledge of $\sigma$, and is then marginalised over. We have not made use of 
this parameter in this work.}. Strictly speaking the noise in the 
\emph{Eddington} case is Poisson distributed (being photon shot noise), 
however given the large number of photons in each time bin used for the 
transit search  this is indistinguishable from a Gaussian noise distribution. 
The likelihood is therefore given by:

\begin{equation} 
\label{liki}
p(data|P,d,o) = \displaystyle{\prod_{i=1}^{N}}\left[
  \frac{\sigma^{-1}}{\sqrt{2\pi}} \times
  \exp\left[-\frac{\left(y_i-y(t_i)\right)^2}{2\sigma^{2}} \right]
\right]
\end{equation}

\noindent where $N$ is the total number of data points.

Re-expressed in terms of the $n+1$ bins of the model:

\begin{equation} 
\label{likj}
  \begin{array}{lcl}
   p(data|P,d,o) & = & \sigma^{-N} (2\pi)^{-N/2} \\
   & \times & \displaystyle{\prod_{j=0}^{n}} \exp\left[-
     \left(\displaystyle{\sum_{i=1}^{n_{j}}}y_{i} -r_{j} \right) ^2 /
          2\sigma^{2} \right]
  \end{array}
\end{equation}

\noindent where $n_{j}$ is the number of data points in bin $j$ and $r_j$ is 
the model value in bin $j$.

As shown in \citet{gre99} the argument of the exponential can be reduced to:

\begin{equation}
\label{eqargexp}
  \frac{ \left( \displaystyle{\sum_{i=1}^{n_{j}}}y_{i} -r_{j} \right) ^2}
          {2\sigma^{2}} = 
  \frac{ W_{j}\left(r_{j}-\overline{d_{W_{j}}}\right)^2+\chi^{2}_{W_{j}}}{2}
\end{equation}
  
This allows the marginalization over the $r_{j}$'s to be performed,
which we do identically to Gregory, to obtain:

\begin{equation}
  \label{eqpdpdo}
  \begin{array}{lcl}
    p(data|P,d,o) & = & \sigma^{-N} (2\pi)^{-N/2} 
      \left(\Delta_{r}\right)^{-(n+1)}
      \left(\frac{\pi}{2}\right)^{(n+1)/2} \\
    & & \\
    & \times & \exp\left[-\displaystyle{\sum_{j=0}^{n}}
                                      \chi^{2}_{W_{j}}/2\right] \\
    & & \\
    & \times &  \displaystyle{\prod_{j=0}^{n}} W_{j}^{1/2} 
                \left[ \textrm{erfc} (y_{j,\mathrm{min}}) - 
                \textrm{erfc} (y_{j,\mathrm{max}}) \right]
   \end{array}
\end{equation}
\noindent where:

\begin{itemize}

\item $\Delta_{r} = r_{\mathrm{max}} - r_{\mathrm{min}}$ is the range of 
  values the model step function is allowed to take;

\item the quantities $W_{j}$, $\chi^{2}_{W_{j}}$, $y_{j,\mathrm{min}}$ and
  $y_{j,\mathrm{max}}$ are taken directly from Eqs.~(11) to (16) in 
  \citet{gre99} ;

\item $\textrm{erfc}(y)$ is the complementary error function.

\end{itemize}

\subsection{Odds ratio calculation}
\label{odds}

In order to use the likelihood to determine a given parameter all the
other parameters must be marginalized, by multiplying by the corresponding 
prior and integrating over the parameter's range of values.  When
marginalizing the phase, in order to minimise the computing time, we
incremented the phase by steps of $\pi d / P$ where $P$ is the
period and $d$ the duration of the transit, corresponding to time offset 
increments of $d/2$.  We used the same priors for each parameter as Gregory, 
with a flat prior for the new parameter $d$. Although we worked in terms of
period rather than frequency this does not change the calculations.

Odds ratios $O_{P,c}$ were then computed by comparing the
probabilities as defined in Eq.~(\ref{eqpdpdo}) for a range of
period values, integrating out all other parameters, to the
probability obtained with a constant model denoted by the subscript
$c$. These odds ratios can then used to check for evidence of a periodic 
signal over the entire frequency range before proceeding to determine 
individual parameters, as described in \citet{gre99}. The a posteriori 
probabilities needed for parameter estimation can be directly evaluated from
the odds ratios by multiplying by the relevant prior and normalizing.

In \citet{gl92b}, a global odds ratio is calculated for each light curve by 
marginalising over all the parameters, in order to determine whether there is 
evidence for a periodic signal. If the global odds ratio is larger than $1$, 
the answer is yes. In that case, posterior probability distributions for 
individual parameters are used to determine the optimal parameter values.

\subsection{Weighting factor to compensate for uneven distribution
  into the bins}
\label{weight}

When the number of periods is low such that one bin might be
represented four times while another only three times, or if there are
gaps in the data which may not be evenly distributed over the bins, 
\citet{gl92b} noted that some of their initial assumptions may fail, leading
to the appearance of an erroneous trend in the posterior probability for the 
period.

In an appendix to \citet{gl92b}, a solution to this problem
was proposed. A weighting factor $s_{j}$ is applied to each bin:

\begin{equation}
\label{eqsj1}
s_{j} = \left( \frac{n_{j} m}{N} \right) ^ {-n_{j}}
\end{equation}

This factor is derived in the context of Poisson statistics and does not 
apply to the present, Gaussian noise case.

Despite the low number of periods in our light curves it was found
that no weighting factor was required in the benchmark algorithm that
reproduced the GL identically.
However it is clear that the problem is more acute in the modified
algorithm. The `out of transit' bin contains many more data points than the 
others, and therefore has a much larger effective weight. A weighting factor 
is required to compensate for this problem. The expression given above for 
$s_{j}$ is only appropriate in the photon count context in which it was 
derived, not in the Gaussian noise case adopted here. A different weighting 
factor can be heuristically derived by considering Eq.~(\ref{eqargexp}). 
The contribution of each model level to the likelihood is a $\chi^2$ sum. The 
variance of a $\chi^2$ distribution is given by the number of degrees of 
freedom $\nu$. In each bin there are $n_j$ data points and $\mathrm{n_{param}}$
parameters to adjust. As $n_j \gg \mathrm{n}_{\mathrm{param}}$, 
$\nu = n_j - \mathrm{n}_{\mathrm{param}} \simeq n_j$.
Weighting each bin by a factor $1/n_j$ is therefore equivalent to weighting 
by the variance. In practice this is achieved by maintaining the expressions 
for $\overline{d_{W_{j}}}$ and $\overline{d^{2}_{W_{j}}}$, given in
 \citet{gre99} in terms of $d_{i}$ and $\sigma$, but replacing $W_{j}$ by 
$W_{j}/n_{j}$.

This modification was implemented in our algorithm and found
to give more robust results.

\subsection{Minimizing the computing time}
\label{compt}

For a given set of parameters, the calculation of the likelihood involves 
summing over each element in each bin. The time required to compute the 
likelihood for a given set of $P$, $d$, $o$ therefore scales linearly 
with the number of points in the light curve. It also increases with the 
number of bins, but this is a slow increase. It does not depend on the 
individual parameter values.

The overall computing time also depends, of course, on how tightly the 
parameter space is sampled. It is necessary to minimise the number of trial 
values for each parameter without missing potentially localised likelihood 
maxima. Because of the relative sharpness of the peak in the posterior
probability for the period, the period increment needs to be kept
fairly small (typically once or twice the time step between data points).
Attention was therefore concentrated on what increment was suitable in
terms of phase. The results are not significantly worsened by
increasing the \emph{posital} phase increment from $1/P$ (i.e. shifting the 
model by 1 sampling time at each increment) to $d/2nP$ (i.e. shifting the model
by half the duration of an in-transit bin at each increment). Further 
increase leads to sharp steps in the posterior
probability distribution (analogous to Shannon's sampling theorem). 

However, the computing time is inversely
proportional to the increment, and the steps in the distribution are
effectively removed by dividing it by the equivalent distribution for
an entirely flat light curve with the same duration, sampling and data gaps as
the light curve. We call this dividing function the ``window function''
\footnote{This also has the advantage of ironing out any residual effects of 
the uneven bin duration not removed by the weighting factor.}. 
We therefore used a phase increment of $d/2P$ and
performed the division before analyzing the results. As the window function 
only needs to be calculated once per set of parameters, this is
much faster than using a smaller increment (see Sect.~\ref{perf}).

Note that due to the use of this window function one should not strictly 
speaking use the word 'posterior probability' when talking about the output 
of the algorithm. In the rest of this paper we  will refer to `modified 
posterior probability' to mean `posterior probability distribution divided by 
the window function'. This also implies that the global odds ratios mentioned 
in Sect.~\ref{odds} cannot be used to directly measure the ratio of the 
probabilities for a periodic model compared to a constant model. Instead, we 
use bootstrap simulations (see Sect.~\ref{perfmet}) to set a threshold 
value of the detection statistic above which a detection is accepted. 

%__________________________________________________________________

\section{Comparing the modified algorithm with the original version}
\label{comp}

\begin{figure}[!tbp]
  \begin{center} 
    \leavevmode 
    \epsfig{file=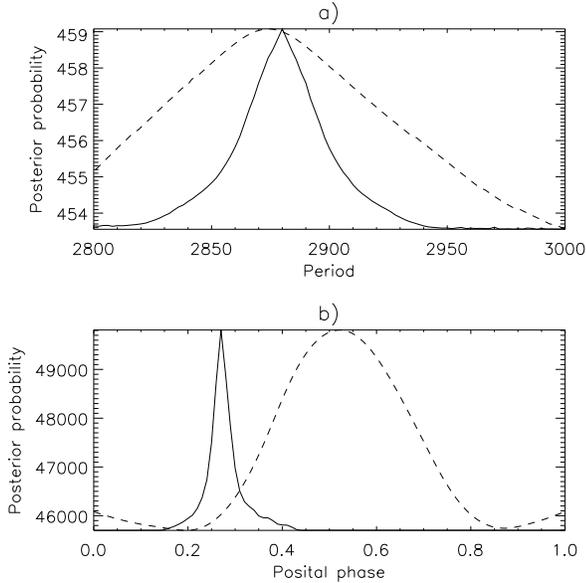,
    width=8.0cm, angle=0} 
  \caption{Comparison of the Gregory-Loredo (GL) and modified methods for 
  the case of Jovian planet transiting accross a $10^{\rm{th}}$ magnitude star 
  (as described in Sect.~\ref{afewcases}), with a period of 2800 
  $\times$ 15 minutes \textbf{(a)} and a phase of 0.25 \textbf{(b)}. 
  Solid line: modified algorithm, dashed line: GL method. Both methods 
  successfully detect the period of the transits although the peak 
  is sharper with the modified method. The GL method is unsuccessful in the 
  phase domain (the GL phase results are folded over the 10 bins). Note that 
  the probabilities are in arbitrary units.}
  \label{fig:comp}
  \end{center}
\end{figure}

In order to establish a reference point and to have a preliminary
estimate of the modified algorithm's performance, some tests were run
on both the original and the modified version. Table \ref{tab:par}
summarizes the names and meanings of the various parameters in each method.

\begin{table}[!tbp]
  \begin{tabular}{lll}
    \hline
    \hline
    Symbol & Method & Meaning \\
    \hline
    $P$   & both     & Period of transits \\
    $\mathrm{ph}$ & both     & Posital $[0-1]$ phase of the transits \\
    $m$   & GL       & Number of steps per period \\
    $d$   & modified & Duration of each transit \\
    $n$   & modified & Number of steps inside each transit \\
    \hline
  \end{tabular}
\caption{List of all parameters for the two types of models tested, with the
  symbols used to refer to them.}
\label{tab:par}
\end{table}

\subsection{A few qualitative tests}
\label{afewcases}

From a typical light curve described below a number of parameters were
varied one by one and the odds ratios were plotted as a function of
period and as a function of phase.  The base light curve lasted
$11\,500 \times 15$ minutes (119.8 days), contained a transiting giant planet 
with a period of 30 days, a duration of 15 hours and a posital phase
(i.e.\ phase in radians divided by $2\pi$) of 0.25. The magnitude of the parent
star was $10.0$, which for \emph{Eddington} corresponds to a signal to
noise ratio of roughly $1400$ over 15 minutes, so that the depth of the
transit for the Jupiter-sized planet is 14 times the noise standard
deviation.  It was analyzed with $m=10$ in the case of the GL method,
and $n=4$ in the case of our method\footnote{The possibility of using $n=1$ 
for detection only purposes, then a larger value of $n$ for transit 
reconstruction, will be the subject of investigations in a further 
paper.\label{fn:n=1}}. In order to sample the transit as well with the GL
method as with the modified method, a much higher value of $m$ would need to 
be used, but this would be too computationally expensive. Instead the values 
of $m$ and $n$ we chosen such that the computing times were similar.
The results obtained for this benchmark case are shown in Fig. 
\ref{fig:comp}. 

Each of the parameters (be they associated with the light curve or with the 
model) was varied over a small range of representative values.
These one-off tests on a small parameter space confirmed some expected
trends:
\begin{itemize}

\item for a given light curve duration the detection is less precise
  for longer periods as the light curve contains less transits;
  
\item as expected, the unmodified GL method is not well suited to
  detecting the phase as there is no way of labeling one bin the first
  one. A detection is still possible by folding the posterior
  probability for the phase over the number of bins used. On the other
  hand the phase is very successfully recovered with the modified
  version, and the precision does not vary with the phase itself;

\item the larger the value of $m$ (GL method), the sharper the
  detection.  However $m=10$ appeared sufficient for our purposes;

\item increasing the value of $n$ (modified method) does not
  necessarily improve the detection ability since one starts to fit
  the noise inside the transits, which is not periodic. When fitting
  Gaussian profiles it is standard to require a minimum of $2$ bins
  per FWHM.  The shape of the transit is not Gaussian but it is
  relatively simple, hence we multiplied by a safety factor of $2$,
  leading to $n=4$ in further calculations. However when dealing with
  a particular value of $d$ it is advantageous to choose $n$ so $d$ is
  a multiple of it to avoid introducing extra noise by splitting individual 
  data points across bin boundaries(see footnote \ref{fn:n=1});

\item although the modified method should in principle allow us to
  determine the duration of the transit, in practice this is not
  successful.  The program may be fitting a much wider region than the
  transit itself. In the GL method, as there are only 10 to 20 bins per
  period, with $P$ of order several hundred sampling times or more, the bin 
  in which the transit falls is much larger than the transit itself. We have 
  seen that the 
  loss of information this implies does not prevent the detection of the 
  period by the GL method. The modified algorithm is likely to overestimate 
  the transit duration because fitting a region larger than the transit does 
  not significantly reduce the likelihood. For now the duration of the 
  transit was simply marginalized
  over; once the presence of a transit is asserted and its period
  known, phase folding should allow a fairly quick determination of
  the shape and duration;

\item for a given set of parameters, with $m=10$ and $n=4$, such that
  both algorithms have similar computing times, the detection peaks
  are much sharper with the modified version.

\end{itemize}

%__________________________________________________________________

\section{Performance of the algorithm}
\label{perf}

\subsection{Method}
\label{perfmet}

To evaluate the performance of the algorithm, we used the same method as 
\citet{ddk+2000}. For each set of trial parameters the algorithm was run 
first on a set of one hundred simulated light curves containing only Gaussian 
noise and no transits. Subsequently it was run on another set of one hundred
simulated light curves containing Jovian-type planetary transits with the 
characteristics described in Sect.~\ref{trans}, with the same level but different 
realizations of the photon noise, and with uniformly distributed random phases

\begin{figure}[!htbp]
  \begin{center} 
    \leavevmode 
    \epsfig{file=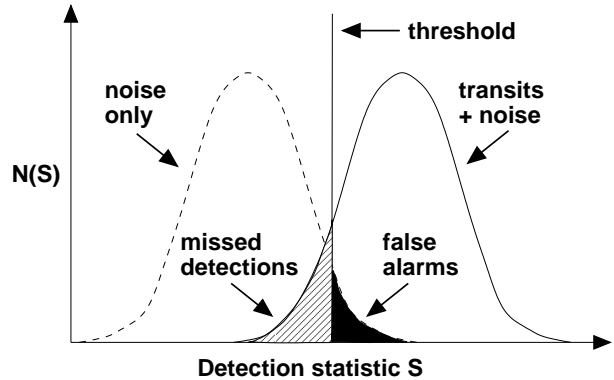,
    width=8.0cm, angle=0} 
  \caption{Schematic diagram of the method used to set the optimal threshold 
    and compute the false alarm and missed detection rate.
     Solid line: Distribution of the detection statistics obtained for 
     lightcurves with noise + transit. 
     Dashed line: Distribution of the detection statistics obtained for 
     lightcurves with noise only.
     Vertical solid line: threshold.
     The hashed area, to the left of the threshold but under the `transits' 
     distribution, corresponds to the missed detection rate. 
     The filled area, to the right of the threshold but under the `noise only'
     distribution, corresponds to the false alarm rate.
   }
  \label{fig:twodist}
  \end{center}
\end{figure}

For each simulation, the modified posterior probabilities were plotted 
versus period and the value of the maximum was noted. This maximum is our 
`detection statistic', on the basis of which we determine whether there is a 
transit or not. We then plot a histogram of the detection statistics measured 
from running the algorithm over all the light curves with transits and one 
histogram for all the light curves with noise only. In other words, one 
histogram corresponds to the cases where the transit hypothesis is correct 
and one to the cases where the null hypothesis is correct. 
Ideally, the two distributions
would be completely separated, with no overlap, and choosing a
detection threshold located between the two histograms would guarantee
a 100\% detection rate and a 0\% false alarm rate. In practice, for
the cases of real interest, close to the noise level, the two
histograms will show an overlap. A compromise has to be found by
choosing a threshold which minimises a penalty factor designed to take into 
account both false alarm and missed detection rate. This is illustrated in 
Fig.~\ref{fig:twodist}. 

Depending on the circumstances, it may be more important to minimise the false 
alarm rate than the missed detection rate. This is the approach followed by 
\citet{jcb02}, on the basis that detections from space experiments are hard 
to follow-up from the ground. An alternative view is any real transit that is 
rejected is a loss of valuable scientific information. As long as the 
false alarm rate is kept to a manageable level, further analysis of the
light curves will prune out the false events. We have opted here for 
an intermediate position, and our penalty factor is simply the sum of the 
missed detection rate $\mathrm{N}_{\mathrm{MD}}$ and the false alarm rate 
$\mathrm{N}_{\mathrm{FA}}$:
\begin{equation}
  \label{eq:penalty}
  \mathrm{F}_{\mathrm{penalty}} = \mathrm{N}_{\mathrm{FA}} + 
                                  \mathrm{N}_{\mathrm{MD}}
\end{equation}

However, the marginalised detection algorithm yields modified posterior 
probabilities as a function of
period, and also as a function of phase. The simultaneous use of
the two detection statistics $\mathrm{S_{per}}$ and $\mathrm{S_{ph}}$ 
(plotting 2-D rather than 1-D distributions) 
increases the discriminating power of the algorithm, (as long as the two 
distributions do not have secondary maxima in 2-D space). This is shown when 
comparing the false alarm and missed detection rates obtained from period and 
phase information separately and together. The threshold in the 2-D case 
takes the form of a line: $\mathrm{S_{ph}} = a + b \times \mathrm{S_{per}}$. 
Here the optimal
values of $a$ and $b$ were found by trial and error, although standard 
discriminant analysis techniques can be used to determine them automatically.

\subsection{Results}
\label{perfres}

\subsubsection{An ideal case}
\label{perfcomf}

In \citet{dbd2001}, analysis performed on the basis of 200 bootstrap 
samples for the COROT observations of a star with magnitude 13 and an 
Earth-sized planet showed, 
with 6 transits lasting 5 hr each, a probability of true detection of around
0.3. We performed the simulations described in Sect.~\ref{perfmet} for a 
similar case: Earth-sized planet orbiting a K5V type star with $V=13$ with a
period of $932 \times 15$ minutes and a transit duration of 5 hr. The light 
curve is sampled with 15 minute bins. The noise is different from the COROT 
case, as we concentrate uniquely on the photon noise expected for 
\emph{Eddington}.

\begin{figure}[!tbp]
  \begin{center} 
    \leavevmode 
    \epsfig{file=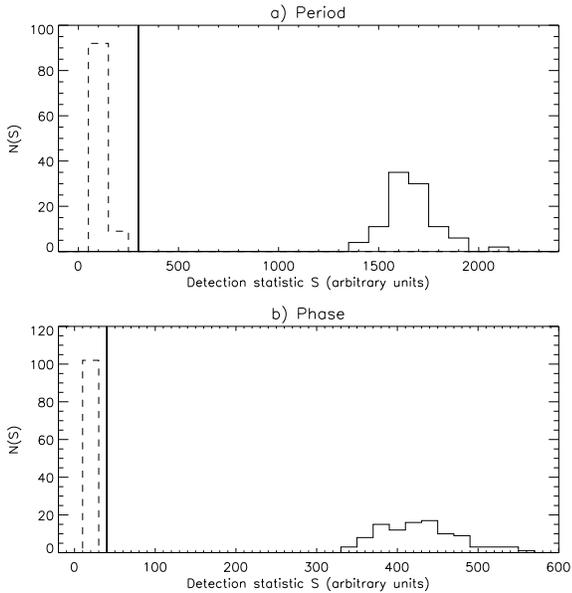,
    width=8.0cm, angle=0} 
  \caption{Distributions of the detection statistics for an Earth-sized 
    planet orbiting a $V=13$ star with period $P=932\times 15 $ minutes. 
     Solid line: lightcurves with noise + transit. 
     Dashed line: lightcurves with noise only.
     Vertical solid line: threshold value.
     \textbf{a)} Period.
     \textbf{b)} Phase. 
     Over 100 realizations there were no false alarms and no missed 
     detections.}
  \label{fig:comfy}
  \end{center}
\end{figure}

The results are shown in Fig. \ref{fig:comfy} for period and phase separately.
As the distributions for the noise only and transit light curves are
completely separated, each parameter alone is sufficient to determine a 
threshold ensuring null false alarm and missed detection rates.

\subsubsection{Performance of the algorithm at the noise limit}
\label{perflim}

Given that the key scientific goal of \emph{Eddington} in the field of
planet-finding is the detection of habitable planets, the performance
of the algorithm was extensively tested for habitable planets at (or
close to) the noise limit of \emph{Eddington}. The case of a
Earth-like planet orbiting a K dwarf in a habitable orbit was used as
benchmark. The light curve was simulated for a system with the following 
parameters:

\begin{itemize}

\item the star is a K5 dwarf ($R_* = 0.8$ $R_\odot$) with a range of
  apparent V-band magnitudes $V=14.0, 14.5, 15.0$;

\item planet with radius $R_{\mathrm{p}} = R_{\oplus}$ and a period of
  4 months, orbiting the star at a distance of $0.64$ A.U. (leading to
  a transit duration of $\approx 10.5$ hours);

\item light curve duration of 16 months, containing 4 transits. The
  light curves were sampled every hour.

\end{itemize}

\begin{figure}[!htbp]
  \begin{center} 
    \leavevmode 
    \epsfig{file=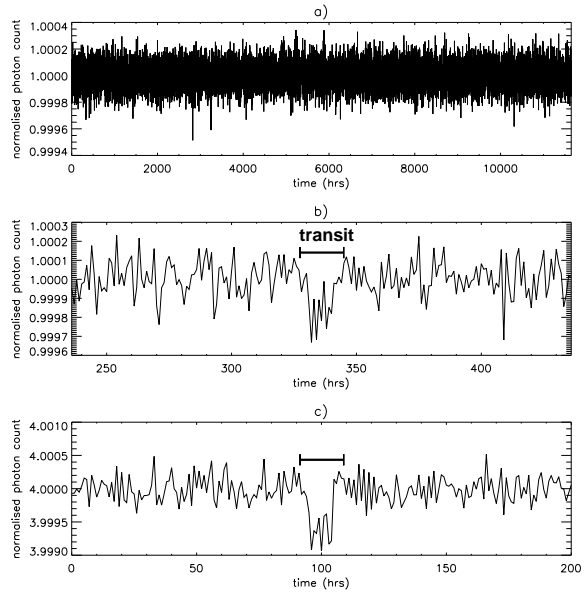,
    width=8.0cm, angle=0} 
  \caption{An example light curve containing 4 transits of an Earth-like 
     planet orbiting a K5V star with $V=14.5$. 
     \textbf{a)} Full light curve.
     \textbf{b)} Portion around a transit.
     \textbf{c)} The four transits phase-folded.}
  \label{fig:exlc}
  \end{center}
\end{figure}

An example of a light curve is shown in Fig. \ref{fig:exlc}.
The resulting transit event has a depth $\Delta F/F = 1.4 \times
10^{-4}$. For the \emph{Eddington} baseline collecting area a star at
$V=14$ will result in a photon count of $1.8 \times 10^8$ per hour, so that 
the Poisson noise standard deviation will be $1.34 \times 10^4$.
The $S/N$ of the transit event in each 1 hour bin will thus be $1.88$. 
Following the same reasoning for the $V=15$ case, the $S/N$ of the 
of transit event in a single one hour bin is $1.19$. As there are 4 transits
lasting 10 hours each in the light curves considered, the overall transit 
signal has a $S/N$ of $\sqrt{40} \times 1.19 \simeq 7.5$.

\begin{figure}[!htbp]
  \begin{center} 
    \leavevmode 
    \epsfig{file=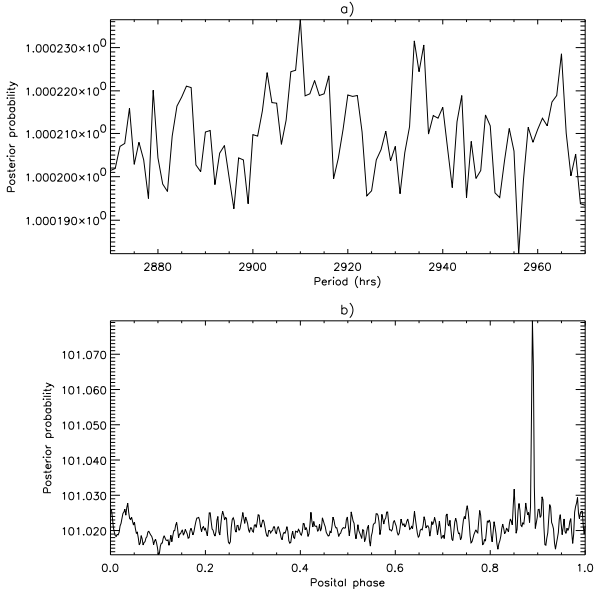,
    width=8.0cm, angle=0} 
  \caption{Example of posterior probability distributions arising from the 
    lightcurve shown in Fig.~\ref{fig:exlc} (arbitrary units).
     \textbf{a)} Period: real value = 2912 hours, error = $-2$ hours.
     \textbf{b)} Phase: real value = 0.885, error = 0.005.}
  \label{fig:exdat}
  \end{center}
\end{figure}

\begin{figure}[!htbp]
  \begin{center} 
    \leavevmode 
    \epsfig{file=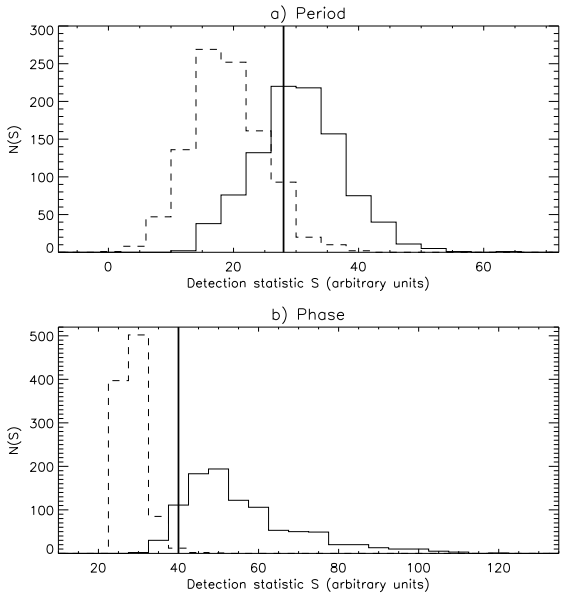,
    width=8.0cm, angle=0} 
  \caption{Distributions of the detection statistics for an Earth-sized 
    planet orbiting a $V=14.5$ star with period $P=4$ months.
     Solid line: lightcurves with noise + transit. 
     Dashed line: lightcurves with noise only.
     Vertical solid line: threshold value.
     \textbf{a)} Period: 190 false alarms and 185 missed detections over 1000
     realizations.
     \textbf{b)} Phase: 27 false alarms and 14 missed detections over 1000 
     realizations.}
  \label{fig:ind}
  \end{center}
\end{figure}

\begin{figure}[!htbp]
  \begin{center} 
    \leavevmode 
    \epsfig{file=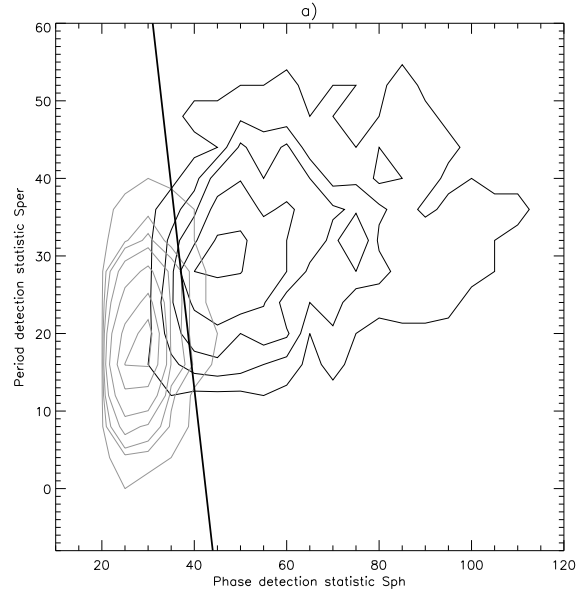,
    width=8.0cm, angle=0} 
    \epsfig{file=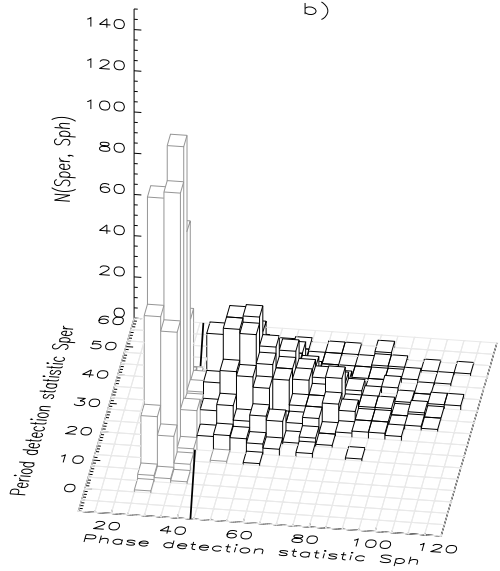,
    width=8.0cm, angle=0} 
  \caption{{\bf a)} Contour plot and {\bf b)} 3-D representation of the 
    two-dimensional distributions of the period and phase detection 
    statistics for an Earth-sized planet orbiting a $V=14.5$ star with 
    period $P=4$ months. 
    Black: lightcurves with noise + transit. 
    Grey: lightcurves with noise only.
    Solid line: Optimal threshold line.
    ($\mathrm{S_{ph}} = 42.47 - 1.191 \times \mathrm{S_{per}}$), yielding 29 
    false alarms and 9 missed detections over 1000 realisations.}
  \label{fig:cont}
  \end{center}
\end{figure}

With the results of the simulations, an example of which is shown in Fig.
\ref{fig:exdat}, the analysis described in Sect.~\ref{perfmet} was performed 
for all three magnitudes, confirming that the combined use of the two 
statistics improves the results. This is illustrated for the $V=14.5$ case in 
Figs.~\ref{fig:ind} \& \ref{fig:cont} (for this particular case 1000 rather 
than 100 runs were computed to improve precision).

\begin{figure}[!htbp]
  \begin{center} 
    \leavevmode 
    \epsfig{file=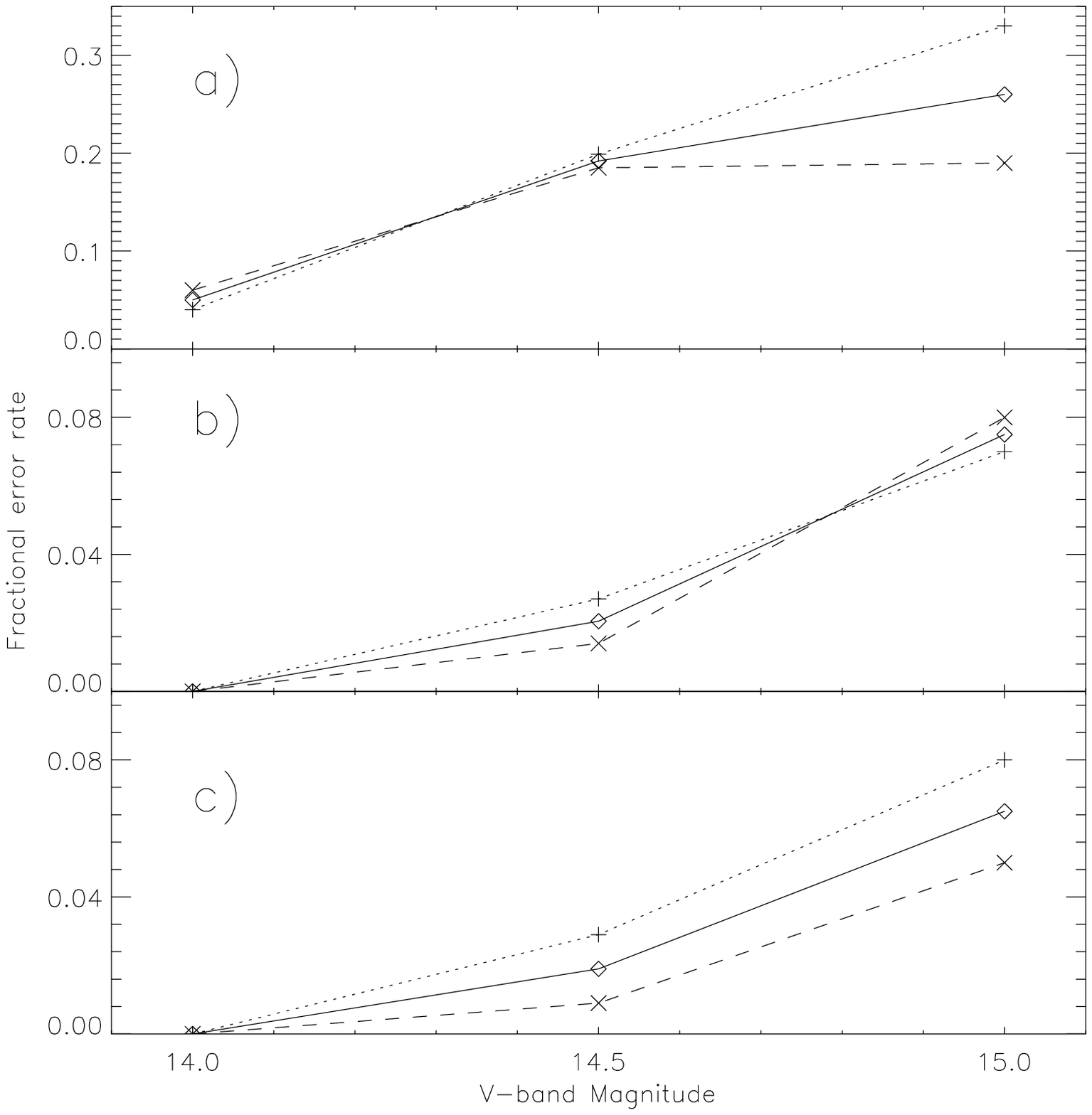,
    width=8.0cm, angle=0} 
  \caption{Evolution of the algorithm's performance (in terms of fractional 
     error rates) with magnitude
     ($P=4$ months, $V=14.5$, Light curve duration $16$ months). 
     \textbf{a)} Using the period statistic only.
     \textbf{b)} Using the phase statistic only
     \textbf{c)} Combining the two statistics
     Dotted line: False alarm rate.
     Dashed line: Missed detection rate.
     Solid line: Mean error rate.}
  \label{fig:perfmag}
  \end{center}
\end{figure}

As illustrated in Fig. \ref{fig:perfmag}, a mean error rate\footnote{i.e.\ the
mean of the false alarm and missed detection rates.} of $<$ 3\% can be 
achieved up to magnitude 14.5. This magnitude is therefore taken as the 
performance limit for the algorithm for an Earth-sized planet around a 
K5V-type star. However this analysis is not complete 
enough to allow a precise determination of the limit, as the noise treatment 
is incomplete (photon noise only being considered) and one would need more 
runs per simulations to compute meaningful errors on the false alarm and 
missed detection rates (sets of 1000 runs, as was done for the
limiting $V=14.5$ case, should be computed for all cases). 

The asymetric shape of the distributions shown in Figs.~\ref{fig:comfy}, 
\ref{fig:ind} \& \ref{fig:cont} implies that, even though the thresholds are 
chosen to minimise false alarms and missed detections equally, the optimal 
threshold results in more false alarms than missed detections. This could 
easily be avoided, if needed, by replacing Eq.~\ref{eq:penalty} by:
\begin{equation}
  \label{eq:penaltymod}
  \mathrm{F}_{\mathrm{penalty}} = A \times \mathrm{N}_{\mathrm{FA}} + 
                                  \mathrm{N}_{\mathrm{MD}}
\end{equation}
\noindent where $A$ is a factor greater than $1$. Alternatively one could 
keep the penalty factor unchanged but set a strict requirement on the maximum 
acceptable false alarm rate.

As in any unbiased search for periodicity in a time-series, the
inclusion of a larger range of periods in the search will lead to a
higher chance of finding a spurious (noise-induced) period signal in
the data. The simulations used here to assess the algorithm's
performance are based on a search through a relatively small range of
periods. In practice, lacking any a priori knowledge of the possible
periodicity of planetary orbits around the star being observed, one
will have to test a large range of periods, ranging from few days (the
physical limit of the period of planetary orbits) all the way to the
duration of the data set (searching for individual transit events).

\subsubsection{Data gaps}
\label{perfgaps}

Any realistic data set will suffer from gaps in the data. While the
orbits of both \emph{Eddington} and \emph{Kepler} have been chosen
to minimize gaps, 100\% availability is not
realistic, and gaps will be present due to e.g.\ telemetry dropouts,
spacecraft momentum dumping maneuvers, showers of solar protons during
large solar flares, etc. For this reason any realistic algorithm must
be robust against the presence gaps in the data, showing graceful
degradation as a function of the fraction of data missing from the
time series. 

We have therefore tested the algorithm discussed here using simulated
light curves with 5\%, 10\% and 20\% data gaps, randomly distributed
in the data, i.e.\ 5\% of the points in the time series are selected randomly
with a uniform distribution and removed from the light curve. The gaps will 
probably not be randomly distributed in reality, but as the typical gap 
duration is expected to be of order 1 or 2 hours, simulated random gaps can 
already be used to test the algorithm's robustness. For reasons of 
computing time, to avoid having to recalculate the ``window function'' at 
each run, the distribution of the data gaps is the same for all runs of a 
simulation. As the gaps are chosen one by one there are rarely gaps of more 
than two consecutive time steps, i.e.\ 2 hours. 
Note that e.g.\ the \emph{Eddington} mission is
designed to produce light curves with a duty cycle $\ge 90\%$, so that
the case with 20\% data gaps represents a worst case analysis. 

\begin{figure}[!tbp]
  \begin{center} 
    \leavevmode 
    \epsfig{file=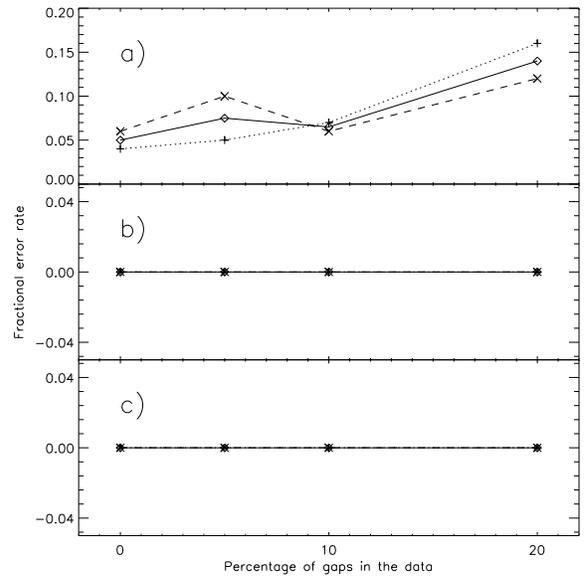,
    width=8.0cm, angle=0} 
  \caption{Evolution of the algorithm's performance with data gaps
     ($P=4$ months, $V=14.0$, Light curve duration $16$ months).
     \textbf{a)} Using the period statistic only.
     \textbf{b)} Using the phase statistic only.
     \textbf{c)} Combining the two statistics.
     Dotted line: False alarm rate.
     Dashed line: Missed detection rate.
     Solid line: Mean error rate.}
  \label{fig:perfgap}
  \end{center}
\end{figure}

The results are shown in Fig. \ref{fig:perfgap}. There is visibly very little 
degradation up to 20\% data gaps. When using $\mathrm{S_{per}}$ alone or the 
two statistics combined
there is no perceptible difference. We can therefore say this this algorithm 
is robust at least for data gaps of the type likely to occur due to 
e.g.\ telemetry dropouts, which last only a few hours. One would also expect 
the algorithm to perform well in the presence of longer gaps: the effect of
gaps is to render the number of samples per bin uneven, and this is already
the case for this particular method with no gaps at all.

\subsubsection{Number of transits in the data}
\label{perfno}

The planetary transits detection phase of the \emph{Eddington} mission is 
planned to last 3 years with a single pointing for the entire duration of 
that phase. There will therefore be three or four transits in the light curve
for a typical habitable planet. However, other missions such as COROT are 
planned with shorter (5 months) pointings and its is of interest for this type 
of mission to study the degradation of the algorithm's performance as the 
number of transits in the light curve reduces. If the algorithm performs well
with 2 or less transits, in the context of \emph{Eddington} it may also allow
the detection of ``cool Jupiters'', i.e.\ Jupiter-sized planets with orbits 
more similar to those of the gaseous giants in our solar system. This would 
be of relevance to the question of how typical our solar system is.

\begin{figure}[!tbp]
  \begin{center} 
    \leavevmode 
    \epsfig{file=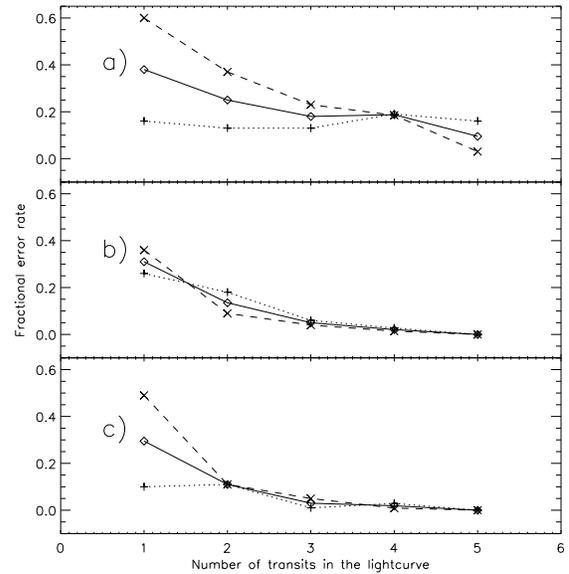,
    width=8.0cm, angle=0} 
  \caption{Evolution of the algorithm's performance with the number of 
     transits in the light curve, i.e.\ the light curve duration
     ($P=4$ months, $V=14.5$). 
     \textbf{a)} Using the period statistic only.
     \textbf{b)} Using the phase statistic only.
     \textbf{c)} Combining the two statistics.
     Dotted line: False alarm rate.
     Dashed line: Missed detection rate.
     Solid line: Mean error rate.}
  \label{fig:perfntr}
  \end{center}
\end{figure}

Sets of 100 runs with the characteristics specified in Sect.~\ref{perflim} 
for a star of magnitude 14.5 were computed for light curve durations of 
4, 8, 12, 16 and 20 months, containing between 1 and 5 transits. The results 
are shown in Fig. \ref{fig:perfntr}. The degradation only becomes significant
when less than three transits are present. However, even mono-transits could
be detectable for larger planets at that magnitude.

\citet{dbd2001} compared a matched filter approach with a Bayesian method 
based on the decomposition of the light curve into its Fourier coefficients.
Their results suggest that the performance degradation in the low number of 
transits case is faster for the Bayesian method than for the matched filter. 
This is because the matched filter makes use of assumptions about the transit 
shape. It is also shown that when the Bayesian method fails to detect 
a transit, it can still reconstruct it if the detection is performed using a 
matched filter. Our algorithm has not been directly compared to a matched 
filter. Its very design is based on the search for a short periodic signal in 
an otherwise flat lightcurve, which is itself an assumption about the shape of 
the signal. The matched filter makes use of more detailed knowledge 
of the transit shape and is therefore likely to perform better in the low 
transit number limit. However our algorithm with $n=1$ may provide already a 
very good approximation to the relatively simple shape that is a transit, and 
therefore perform nearly as well.

\subsubsection{Differences in the two statistics}
\label{2stats}

The two a posteriori probabilities show a different behavior. In general
the phase statistic is far more discriminatory than the period statistic.
The period statistic's lesser effectiveness may be explained in the following 
way. If the phase is wrong, even if the period is right, it is likely 
none of the transits will be matched. If the phase is right, whatever the 
period, at least the first transit will be matched by the model. First we 
consider the likelihood distribution a function of phase, normalised over 
all periods. For an incorrect phase the contribution from the correct period 
is nil as all transits are missed, but for the correct phase all trial 
periods produce a non-negligible contribution (the correct period of course 
contributing most). The likelihood distribution as a function of phase is 
therefore sharply peaked. Then we consider the likelihood distribution as a 
function of period, normalised over all phases. The contribution from the 
correct phase is non-negligible whatever the period. When the period is 
correct, the contribution from the correct phase is washed out by the
contributions from all the incorrect phases. The likelihood 
distribution as a function of period is therefore less sharply peaked.

However the combined use of the two parameters is more successful than the 
phase statistic alone. The reason for this is illustrated in Fig. 
\ref{fig:cont}: in 2-D space the two distributions are aligned on a diagonal,
such that no single value cutoff is optimal in either direction, compared to 
the line shown. In an upcoming paper, the direct use of a combined statistic 
shall be investigated. 
The global odds ratio described in Sect.~\ref{odds} could be used for such 
a purpose. We have noted in Sect.~\ref{compt} that the global odds ratio for 
a given lightcurve cannot be used as an absolute statitstic in the context of 
the present method. It 
can however be used as relative detection statistic, like $\mathrm{S_{per}}$ 
\& $\mathrm{S_{ph}}$, combined with bootstrap simulations.

%__________________________________________________________________

\section{Discussion}

Efficient data processing is one of the challenges for the upcoming
generation of large scale searches for exo-planets through photometric
transits. While radial velocity searches concentrate on limited
number of stars, transit searches will investigate simultaneously
large numbers of stars, and produce large amounts of data (photometric
light curves) for each of them. A computationally efficient and robust
algorithm for the processing of these data sets is necessary to make
transit searches feasible. It is likely that the photometric time
series which represent the observational product of the transit
searches will be analyzed in different stages, using more than a
single approach. In particular, a first level of processing (after
instrumental effects have been removed) should concentrate on singling
out high-probability transit candidates, while efficiently pruning out
the large number (more than 90\%, even if all stars have planets, due
to the low probability of transit events) of light curves in which no
transits are present. In this first stage of analysis the ability to
efficiently screen real transits in the data -- even at the price of a
moderate number of false alarms -- is a key requirement for the
algorithm. The candidate light curves in which a transit is suspected
will then later be subject to a more detailed processing, which can
then afford to be computationally less efficient (given it has to
operate on a much smaller amount of data).

The algorithm we have developed and discussed here is able 
to detect transit events at the limit of the photon noise
present in the light curve. It shows a graceful degradation
of its performance as function of different parameters of interest,
e.g.\ the noise level in the data, as well as the presence of data
gaps and the number of transits actually observed. Its strong
sensitivity to the phase of periodic transits supplies significant
additional information to be then used by further steps of processing
for e.g.\ the reconstruction of the transit parameters. Thus, while
little used in astronomy, Bayesian algorithms appear to be a powerful
tool in the processing of transit data.

\section{Conclusions and future work}
\label{concl}

A novel algorithm to detect transits due to extra-solar planets in
stellar light curves has been developed and tested. The algorithm,
based on a Bayesian approach, has proved successful in the tests
performed so far, which include the effects of photon noise and data
gaps. Using the photometric accuracy and throughput expected for the
\emph{Eddington} mission, we are able to detect an Earth-sized planet orbiting 
a K5V-type star with a period of 4 months down to an apparent stellar 
magnitude of $V \simeq 14.5$.  Randomly distributed data
gaps lasting up to two hours each and covering up to 20\% of the light
curve do not significantly affect the performance of the algorithm.
The minimum number of transits in one light curve required for high
confidence detections is three, however the algorithm's performance
degrades gracefully for small number of transits, so that detections
are possible for individual transits, albeit at a lower confidence
level. This will allow for the detection of larger planets in
long-period orbits (analogous to the gaseous giants of our solar
system), likely to transit only once in the three year planet
detection phase planned for the \emph{Eddington} mission.

The most serious additional noise source to perturb
planetary transit detections from space, is likely to be intrinsic stellar
micro-variability (mostly activity-induced). At the moment it is also the 
least well investigated. The consequences of
activity on the detection efficiency (using simulated light curves
based on the solar light curves recorded by the VIRGO instrument on
board SOHO, which spans all solar activity levels, from solar minimum
to solar maximum) will be the subject of a future paper, in which
the feasibility and effectiveness of using color information, as well
as a number of pre-processing techniques such as whitening, will also be
investigated.

The algorithm we have developed and discussed here has the potential to form 
part of a powerfull, multi-stage approach to analysing transit lightcurves. 
A more optimised processing method will be discussed in a separate paper.
It will include a variability filtering stage, followed by distinct 
detection and parameter estimation stages, using a combination of a matched 
filter approach and of the present algorithm. 

The performance of the algorithm presented here shows that the search
of planetary transits with amplitudes comparable to the intrinsic
noise level of the data set is fully feasible, and thus represents an
important element in the development of the future generation of
transit-based planet finding missions. 

%______________________________________________________________

\begin{acknowledgements}

  This work was supported by the European Space Agency's Young
  Graduate Trainee program.  The authors are grateful to Hans Deeg for making 
  his transit modelling package UTM available. We wish to thank Mike 
  Irwin, David Katz \& Rolf Jansen for advice, and Celine Defa\"{y} for advice 
  and access to her algorithm. We are also grateful for the thorough and 
  helpful comments of the referee.

\end{acknowledgements}

\bibliographystyle{aa}
\bibliography{h3360}

\end{document}